\documentclass[showpacs, nofootinbib, superscriptaddress, twocolumn,preprintnumbers,amsmath,amssymb]{revtex4-1}

\usepackage{graphicx}
\usepackage{dcolumn}
\usepackage{bm}
\usepackage{color}

\newcommand{\bq}{\begin{equation}}
\newcommand{\eq}{\end{equation}}
\newcommand{\bqa}{\begin{eqnarray}}
\newcommand{\eqa}{\end{eqnarray}}
\newcommand{\nn}{\nonumber \\}

\newcommand\p{\partial}

\newcommand\lm{{\lambda}}
\newcommand\ep{{\epsilon}}

\newcommand\di{{\rm dim}}
\newcommand\ke{{\rm ker}}
\newcommand\cok{{\rm coker}}

\def\be     {\begin{equation}}
\def\ee     {\end{equation}}
\def\bea        {\begin{eqnarray}}
\def\eea        {\end{eqnarray}}
\def\bnn    {\begin{eqnarray*}}
\def\enn    {\end{eqnarray*}}

\begin{document}

\title{
Higher-form symmetries and $d-wave$ superconductors from doped Mott insulators
}

\author{Ki-Seok Kim}

\affiliation{Department of Physics, POSTECH, Pohang, Gyeongbuk 37673, Korea}

\author{Yuji Hirono}

\affiliation{Asia Pacific Center for Theoretical Physics (APCTP), Pohang, Gyeongbuk 37673, Korea}
\affiliation{Department of Physics, POSTECH, Pohang, Gyeongbuk 37673, Korea}

\date{\today}

\begin{abstract}
One path to high-temperature cuprate superconductors is
doping a Mott insulator.
In this paper, we study this system from the view point of higher-form symmetries.
On the introduction of slave bosons, the $t-J$ model at a finite hole doping
can be written in the form of $U(1)$ gauge theories.
After a duality transformation,
they can be written as a generalized $BF$ theory,
in which the higher-form symmetries are more manifest.
We identify the emergent continuous and discrete higher-form symmetries in both $s-wave$ and $d-wave$ superconducting phases, expected to be realized from a doped Mott insulator.
The existence of a topological order is tested by
examining if there is a spontaneously broken discrete one-form symmetry.
We claim that a spontaneous breaking of discrete one-form symmetry may extend to a phase that has massless Dirac fermions in the limit of large number of flavors.
We discuss the  possibility of
a topological phase transition inside the superconducting dome of high $T_{c}$ cuprates.
\end{abstract}


\maketitle


\section{Introduction}

Topological orders provide us with a finer classification of gapped quantum phases beyond the Landau-Ginzburg theory based on symmetry breaking patterns \cite{Wen:1989iv}.
Topologically ordered states exhibit such properties as fractionalized topological excitations, long-range entanglement, and degeneracy of the ground states that depend on the spacetime topology \cite{Wen:2004ym, Kitaev:2005dm}.
A conventional $s-wave$ superconductor is
an example of topologically ordered phase \cite{Hansson:2004wca},
where the key ingredient is that the Cooper pairs have charge $2e$.
Because of this, if we compactify one spatial direction,
we can insert a half magnetic flux in this direction.
This state has exactly the same energy as the state with no flux,
and the ground states become doubly degenerate.
Since the superconductivity does not involve symmetry breaking ($U(1)$ gauge symmetry is not a symmetry), the distinction of a superconducting state with a normal state is given by the existence of a topological order.

In certain cases, the appearance of topological order can be understood as a consequence of a spontaneous breaking of higher-form symmetries \cite{NUSSINOV2009977, Nussinov:2006iva, gaiotto2015generalized, Wen:2018zux},
which is a generalization of symmetry.
The charged objects of higher-form symmetries are extended objects like line operators or surface operators, and
an ordinary symmetry corresponds to a zero-form symmetry, where the corresponding charged object is point-like.
The concepts such as Noether theorem and Nambu-Goldstone (NG) theorem are also generalized to higher-form symmetries.
For example, photons can be understood as NG modes associated with the spontaneous breaking of continuous one-form symmetry. 
A topological order appears
when a discrete higher-form symmetry is spontaneously broken.
The example of $s-wave$ superconductivity is in this category.
The effective theory can be written as an Abelian Higgs model, and
after taking an Abelian duality and dropping massive excitations,
it can be written as a topological $BF$ theory.
The action has a pair of emergent $Z_2$ one-form symmetries (in 2+1 dimensions),
and the corresponding charged objects are Wilson loops and vortex operators. 
Those two operators obey fractionalized statistics.
In the large-loop limit, the expectation values of those operators show the perimeter law,
which means that both one-form symmetries are spontaneously broken,
and the system acquires a topological order.

In this paper, we study the structure of higher-form symmetries of
two-dimensional high temperature superconductors.
The realization of superconductivity from doped Mott insulators is
known as a spin-liquid scenario for high $T_{c}$ cuprates \cite{Lee:2006zzc}.
Resorting to the $U(1)$ slave-boson representation of the $t-J$ effective Hamiltonian,
an effective field theory for $d-wave$ superconductors is obtained from a spin liquid state,
based on the assumption of the separation of spin and charge degrees of freedom.
We take an Abelian duality to see the symmetry structure in a manifest way, and
identify the emergent higher-form symmetries in the dual effective theories.
By examining whether those symmetries are spontaneously broken or not,
we clarify the nature of topological order in such systems.
In this analysis, electromagnetic $U(1)$ gauge fields are taken to be background potentials.
As a result, ordinary superconductors are identified with superfluids, that are not topologically ordered.

When massless Dirac fermions are absent, which corresponds to $s-wave$ superconductors from doped Mott insulators,
we find that the superconducting state has a $Z_{5}$ one-form symmetry.
This symmetry turns out to be not spontaneously broken, and
there is no topological order associated with this symmetry.
In fact, the $Z_5$ one-form symmetry is shown to be
a subgroup of a continuous $U(1)$ one-form symmetry of this state.
A generalized version of the Coleman-Mermin-Wagner theorem for the continuous higher-form symmetries prohibits the spontaneous breaking of continuous one-form symmetry in $2+1$ dimensions, and hence the $Z_5$ symmetry cannot be broken either.
And yet, the $Z_5$ symmetry is observable in the braiding phase of quasiparticles, as we show in the main text.
From this observation, we conclude that
the topological property of $s-wave$ superconducting phase from $U(1)$ spin liquids
is the same as conventional $s-wave$ superconductivity from Landau's Fermi liquids \footnote{
A similar situation occurs in the dense QCD matter \cite{Fukushima:2010bq}.
There is a scenario that hadronic superfluid phase can be continuously connected to a color superconducting phase, called quark-hadron continuity \cite{Schafer:1998ef}.
This observation is based on the fact that those two phases have the same (zero-form) symmetries.
The continuity is recently extended to include the topological orders
through the analysis of higher-form symmetries \cite{Hirono:2018fjr, Hirono:2019oup}.
}.

One subtle issue regarding topological order is that there are massless Dirac fermions that statistically interact with vortices \cite{Senthil:1999czm, Franz:2002qy}.
In the presence of such coupling to fermions, the discrete one-form symmetry is absent.
However, we point out that,
if we consider the large$-N_{f}$ (number of fermions) limit of
$d-wave$ superconductors (this phase has massless Dirac fermions),
at which a conformal fixed point is realized,
a $Z_2$ one-form symmetry is effectively restored and further it is spontaneously broken.
As a result, we claim that $d-wave$ superconductors from doped Mott insulators cannot be smoothly connected to those from Landau's Fermi liquids, because they can be distinguished by a discrete one-form symmetry. This is in contrast to the case of $s-wave$ superconductors.

The rest of the paper is organized as follows.
In Sec.~\ref{sec:bf-review}, we give a review of a generalized $BF$ theory
for the description of superfluids with topological order.
In Sec.~\ref{sec:d-wave}, we study the topological properties of
$d-wave$ superconductors realized from Landau's Fermi liquids.
In Sec.~\ref{sec:d-wave-mott}, we study the topological nature of
$d-wave$ superconductors from doped Mott insulators.
In Sec.~\ref{sec:conclusion} we give concluding remarks.

\section{
Generalized $BF$ theory for superfluidity with topological order
}\label{sec:bf-review}

\subsection{Effective gauge theory for superfluids with topological order}

A generalized $BF$-type theory in $3+1$ dimensions is
studied recently
that can describe the situations where topological order and superfluidity coexist \cite{Hirono:2018fjr, Hirono:2019oup}.
The structure of higher-form symmetries is analyzed,
and such analysis is useful in classifying quantum phases,
because a spontaneous breaking of discrete higher-form symmetries results in a topological order.
The formalism is applied to the color-flavor locked phase of color superconductivity
of QCD matter.
Since this mathematical structure is also applicable to the present study,
let us briefly review it here in the case of $2+1$ dimensions.

The starting point is a theory with multiple $U(1)$ symmetries,
in which some parts of them are coupled with gauge fields through a covariant derivative. At low energies, the only relevant degrees of freedom are
the massless NG modes and topological excitations.
Such a situation can be captured with the following Lagrangian
of a generalized Abelian Higgs model,
\begin{widetext}
\bqa && S_{\rm AH}
=
\int d^{3} x \Big\{ \frac{1}{2} [\bm H]_{ff'}
(\partial_{\mu} \phi^{f} + [\bm K]_{fc} a_{\mu}^{c})
(\partial_{\mu} \phi^{f'}+ [\bm K]_{f'c'} a_{\mu}^{c'})
+ \frac{1}{2} [\bm G]_{cc'} A_{\mu\nu}^c A_{\mu\nu}^{c'}
\Big\} , \eqa
\end{widetext}
where $\phi^{f}$ are $2\pi$-periodic scalar fields
representing the phases of Cooper pair fields,
$a^c_\mu$ are $U(1)$ gauge fields,
and
$A_{\mu\nu}^{c} \equiv \p_{\mu} a_{\nu}^{c} - \p_{\nu} a_{\mu}^{c}$
are the field strength tensors for $a^c_\mu$.
There are multiple scalar and gauge fields, and
$f$ and $c$ are the indices of the scalar and gauge fields, respectively.

To study the existence of topological order,
let us take an Abelian duality transformation so that the higher-form symmetries of the system can be seen in a manifest way.
The dual action is written as
\begin{widetext}
\bqa
&& S_{\rm Dual}
= \int d^{3} x \Big\{
\frac{1}{8\pi} [\bm{H}^{-1}]_{ff'}
B_{\mu\nu}^f
B_{\mu\nu}^{f'}
+ \frac{ [\bm K]_{fc}}{2\pi} \ep_{\mu\nu\lm} b_{\mu}^{f} \p_{\nu} a_{\lm}^{c}
+ \frac{1}{2} [\bm G]_{cc'}
A_{\mu\nu}^c A_{\mu\nu}^{c'}
\Big\} ,
 \nn \label{Dual_Field_Theory_CFL}
\eqa
\end{widetext}
where
$b_{\mu}^{f}$ is a one-form gauge field
 and
 $B_{\mu\nu}^{f} = \p_{\mu} b_{\nu}^{f} - \partial_{\nu} b_{\mu}^{f}$
 is a two-form field strength tensor for $b_\mu^f$.
 The field $b_\mu^f$
 is dual to $\phi^f$, which
 describe sound modes (supercurrent fluctuations).
The effective theory is invariant under the gauge transformations,
\bqa
&&
a_{\mu}^{c} \mapsto a_{\mu}^{c} + \p_{\mu} \lm^{c} ,
\quad
 b_{\mu}^{f} \mapsto b_{\mu}^{f} + \p_{\mu} \lm^{f} ,
 \eqa
where $\lambda^c$ and $\lambda^f$ are $2\pi$-periodic parameters.
The gauge fields satisfy the standard Dirac quantization condition,
\be
\int d S_{\mu\nu} A_{\mu\nu}^c \in 2\pi Z,
\quad
\int d S_{\mu\nu} B_{\mu\nu}^f \in 2\pi Z,
\ee
where the integrations are over closed 2-dimensional submanifolds.
The gauge invariance requires that the entries of $[\bm K]_{fc}$ are integers. 

The final action is obtained by neglecting the kinetic terms of the massive sound modes and photons.
Since the number of flavor is not necessarily the same as that of color,
the matrix $\bm{K}$ is non-square in general
\footnote{
Although a matrix $BF$ theory is used in \cite{Hirono:2018fjr, Hirono:2019oup} (which is necessary for $3+1$ dimensions),
in $2+1$ dimensions, it is also possible to treat all the one-form fields on equal footing and write the action in the form of a Chern-Simons theory with matrix coefficients.
When there is no diagonal term such as $\int a \wedge d a$, the $BF$-type description is more economical and we adopt this.
}.
When the $\bm K$ matrix has a nontrivial cokernel, i.e.,
$\dim \, (\cok \, \bm{K}) \neq 0$,
there are  gapless superfluid sound modes.
The number of gapless modes is given by $\dim \, (\cok \, \bm{K})$.
Those massless modes can be identified as
$[\bm{b}_{0}]_{\mu}^{f} = P_{ff'}^{b} [\bm{b}]_{\mu}^{f'}$,
where
$P_{ff'}^{b} \equiv \delta_{ff'} - [\bm{K} \bm{K}^{+}]_{ff'}$
is the orthogonal projection matrix to $\cok \, \bm K$,
and $\bm K^+$ is the Moore-Penrose inverse \cite{wiki:mpinverse} of $\bm K$.
Similarly, when $\bm K$ has a nontrivial kernel, i.e.,
$\di ~ (\ke ~ \bm{K}) \not= 0$, we have gapless photons,
and they are identified as
$[\bm{a}_{0}]_{\mu}^{c} = P_{cc'}^{a} [\bm{a}]_{\mu}^{c'}$,
where $P_{cc'}^{a} = \delta_{cc'} - [\bm{K}^{+} \bm{K}]_{cc'}$
is the projection matrix to $\ke \, \bm{K}$.

\subsection{Higher-form symmetries}

Now we are ready to discuss higher-form symmetries of the dual effective theory.
Equation (\ref{Dual_Field_Theory_CFL}) is
invariant under the following transformation,
\bqa
&&
b_{\mu}^{f} \mapsto
b_{\mu}^{f} + q^T_c [\bm{K}^{+}]_{cf} \lambda_{\mu},
\label{One_Form_Transformation}
\eqa
where
$q_c$ is an integer vector chosen
from the orthogonal complement of $\ke \,\bm K$, i.e.,
$q_{c} \in (\ke \, \bm K)^\perp$,
and
$\lambda_{\mu}$ is a one-form field
that satisfies the curvature-free condition,
\bqa
&&
\partial_{\mu} \lambda_{\nu} - \partial_{\nu} \lambda_{\mu} = 0 ,
\eqa
with the normalization
\bqa && \int_{C} d l_{\mu} \lambda_{\mu} \in 2 \pi Z , \eqa
when integrated over a closed loop $C$.
Under the transformation (\ref{One_Form_Transformation}),
the kinetic-energy term for supercurrent fluctuations
is not affected, since
$\delta [P_{ff'}^{b} b_{\mu}^{f'}] = 0$.
The $BF$ term changes as
\bqa &&
\delta_{1} S_{\rm Dual}
=
\frac{1}{2\pi} q^T_c [\bm{K}^{+} \bm{K}]_{cc'} \int d^{3} x
\epsilon_{\mu\nu\lambda} \lambda_{\mu} \partial_{\nu} a_{\lambda}^{c'} .
\eqa
Noting that $[\bm K^+ \bm K]_{c c'}$ is a orthogonal projection matrix to $ (\ke \, \bm K)^\perp$
and we have chosen $q^T_c \in (\ke \, \bm K)^\perp$,
we have
$q^T_c [\bm K^+ \bm K]_{c c'} = q^T_{c'}$
and
\bqa && \delta_{1} S_{\rm Dual} \in 2 \pi Z . \eqa
Thus, the weight of the path integral is unchanged
under this transformation.
If this transformation acts on physical operators (vortex operators) nontrivially,
this is a symmetry of the system.
The entries of $[\bm{K}^{+}]_{cf}$ are in general fractional numbers,
and in that case a discrete one-form symmetry.
The charged objects under this symmetry are vortex operators,
\bqa && V_{p}(C) = \exp\Big( i p_{f} \int_{C} d l_{\mu} b_{\mu}^{f} \Big) ,
\eqa
where $p_{f}$ is a charge vector and $C$ is a closed one dimensional manifold.
This is transformed by the one-form symmetry as
\bqa && V_{p}(C) \mapsto \exp\Big( 2 \pi i p_{f} [\bm{K}^{+}]_{cf} \Big) V_{p}(C).  \eqa

Similarly, one may consider the following transformation,
\bqa && a_{\mu}^{c} \mapsto
a_{\mu}^{c} - [\bm{K}^{+}]_{cf} p_f  \omega_{\mu} , \eqa
where
$p_f \in (\cok \, \bm K)^\perp$ is an integer vector, and
$\omega_{\mu}$ is a flat one-form field satisfying
the curvature-free condition
$\partial_{\mu} \omega_{\nu} - \partial_{\nu} \omega_{\mu} = 0$
and is normalized as $\int_{C} d l_{\mu} \omega_{\mu} = 2 \pi Z$. Here, $(\cok \, \bm K)^\perp$ indicates the orthogonal complement of $\cok \, \bm K$.
Under this transformation, the action is varied as
\bqa && \delta_{1} S_{\rm Dual} =
- \frac{1}{2\pi} [\bm{K} \bm{K}^{+}]_{ff'} p_f \int d^{3} x \epsilon_{\mu\nu\lambda} \partial_{\mu} b_{\nu}^{f'} \omega_{\lambda} .
\eqa 
Noting that $\bm K \bm K^+$ is a symmetric matrix and
$[\bm{K} \bm{K}^{+}]_{f'f} p_f = p_{f'}$, we have
\bqa && \delta_{1} S_{\rm Dual} \in 2 \pi Z . \eqa
The charged objects under this symmetry are the Wilson loops,
\bqa && W_{q}(C) = \exp\Big( i q_{c} \int_{C} d l_{\mu} a_{\mu}^{c} \Big) , \eqa
characterized by a charge vector $q_c$.

In addition to discrete one-form symmetries,
the effective dual field theory Eq. (\ref{Dual_Field_Theory_CFL})  has the following continuous $U(1)$ one-form symmetry,
\bqa &&
a_{\mu}^{c}
\mapsto
a_{\mu}^{c} + \epsilon [\bm{C}^{\alpha}]_{c} \lambda_{\mu},
\eqa
and
\bqa &&
b_{\mu}^{f}
\mapsto
b_{\mu}^{f} + \epsilon [\bm{D}^{\bar{\alpha}}]_{f} \lambda_{\mu}.
\eqa
where  $\epsilon$ is a continuous parameter, and
$[\bm{C}^{\alpha}]_{c}$ and $[\bm{D}^{\bar{\alpha}}]_{f}$ are
basis vectors of the kernel and cokernel of the $\bm{K}$ matrix,
\bqa && [\bm{K}]_{fc} ~ [\bm{C}^{\alpha}]_{c} = 0 , ~~~~~ [\bm{D}^{\bar{\alpha}}]_{f} ~ [\bm{K}]_{fc} = 0 . \eqa
Those vectors are labeled by $\alpha$ and $\bar{\alpha}$,
\bqa &&
\alpha = 1, ~ ..., ~ \di ~ (\ke \, \bm{K}) ,
\nn && \bar{\alpha} = 1, ~ ..., ~ \di ~ (\cok \, \bm{K}) .
\eqa

\subsection{Statistics of quasi-particles and vortices}

In the current setting, not all the Wilson operators or vortex operators are topological,
because of the presence of massless NG modes.
A topological operator is an operator invariant under the deformation of the underlying manifold,
\be
W_q (C + \delta C) = W_q(C),
\ee
when the deformation does not cross other operators.
This does not hold if the operator couples to massless modes.
Even in that case, it is possible to extract the information about the braiding phase
of those operators.
It can be shown that \cite{Hirono:2019oup}
\bqa
&&
\frac{\langle W_{q}(C) V_{p}(C') \rangle}
{\langle W_{q}(C) \rangle \langle V_{p}(C') \rangle} \nn &&
= \exp\left[- 2 \pi i (q_{c} [\bm{K}^{+}]_{cf} p_{f}) ~ Lk(C,C') \right] ,
\label{braiding_phase}
\eqa
where $Lk(C,C')$ is the Gauss linking number of the
two world lines $C$ and $C'$.
Their mutual statistics is encoded in $q^T_{c} [\bm{K}^{+}]_{cf} p_{f}$.

The braiding phase (\ref{braiding_phase}) is closely related to the
discrete one-form symmetries of the system.
Suppose a Wilson loop associated with a charge vector $q$ is topological.
Then, the loop $C$ can be contracted to a point, which means that
$\langle W_{q}(C) \rangle = 1$ (up to the part obeying the perimeter law).
When $C$ is singly linked to $C'$, we have
\be
\langle W_{q}(C) V_{p}(C') \rangle
=
e^{- 2 \pi i (q_{c} [\bm{K}^{+}]_{cf} p_{f}) } \langle V_{p}(C') \rangle .
\ee
This is nothing but the definition of the existence of a one-form symmetry \cite{gaiotto2015generalized}.
Here, $W_{q} (C)$ is the generator of the symmetry and $V_{p}(C')$ is a charge object.

The existence of (discrete) one-form symmetry is not enough for a system
to be topologically ordered.
There has to be a discrete one-form symmetry which is spontaneously broken,
to have a topological order.
This means that at large loop $C$,
the corresponding charged object of the symmetry should behave as
\be
\langle
V_p (C) \simeq \exp \left(  - \kappa \, {\rm perimeter}[C]\right).
\ee
In this case, the system acquires degeneracy depending on the spacetime topology.
If the vortex couples to massless modes, it decays faster than the perimeter law and the symmetry may not be broken.

The condition for the existence of topological order can be summarized as follows:
There exists a pair of integer vectors $(p, q) \in (\cok \, \bm K)^\perp \times (\ke \, \bm K)^\perp$
such that
\be
e^{ i q^T_c [\bm K^+]_{cf} p_f } \neq 1.
\label{eq:cond-phase}
\ee
This condition can be interpreted as that for the presence of mixed 't~Hooft anomalies of one-form symmetries \cite{Hirono:2019oup}.
One way to see  this is introducing background two-form gauge fields for a pair of (discrete) one-form symmetries and
considering the partition function.
The phase of the partition function may become ambiguous in the presence of those background gauge fields, 
if the factor (\ref{eq:cond-phase}) is different from $1$.
This indicates the existence of a 't~Hooft anomaly between those symmetries.
In such a case, we cannot realize the system in $2+1$ dimensions in a gauge invariant way
and we have to couple it to a symmetry-protected topological (SPT) phase
in $3+1$ dimensions. Then, the 't~Hooft anomaly of the boundary state is cancelled by the anomaly inflow mechanism from the SPT bulk phase \cite{Callan:1984sa}.
Since this anomaly matching structure is invariant under renormalization group,
the presence of the 't~Hooft anomaly precludes
a trivial ground state for the $2+1$-dimensional theory.
In the present case, the anomaly is matched by a topological order in the infrared.

\section{
Topological property of $d-wave$ superconductors from Landau's Fermi liquids
}\label{sec:d-wave}

The dynamics of superfluid fluctuations at low energies
is described by the following effective action
\bqa && S_{\rm s-wave} = \frac{\rho_{cp}}{2} \int_{0}^{\beta} d \tau \int d^{2} x (\partial_{\mu} \phi_{cp})^{2} , \eqa
where $\rho_{cp}$ is a phase stiffness parameter proportional to the Cooper-pair density and $\phi_{cp}$ is a phase field to describe sound modes. Here, we consider two space dimensions in the Euclidean time, where $\beta$ is the inverse of temperature, set to be infinite, i.e., considering zero temperature. Performing the duality transformation, we obtain an effective dual field theory in terms of vortices $\Phi_{cp}$ and superfluid sound modes $b_{\mu}^{cp}$, where the partition function is
\bqa
&& Z_{\rm s-wave} = \int D \Phi_{cp} D b_{\mu}^{cp} \exp\Big( - S_{\rm Dual}^{\rm s-wave} \Big) ,
\eqa
and the dual effective action is
\bqa && S_{\rm Dual}^{\rm s-wave} =
\int_{0}^{\beta} d \tau \int d^{2} x \Big\{ |(\partial_{\mu} - i b_{\mu}^{cp}) \Phi_{cp}|^{2} + m_{cp}^{2} |\Phi_{cp}|^{2} \nn && + \frac{u_{cp}}{2} |\Phi_{cp}|^{4} + \frac{1}{2 \rho_{cp}} (\epsilon_{\mu\nu\lambda} \partial_{\nu} b_{\lambda}^{cp})^{2} \Big\} . \eqa

To describe the physics of $d-wave$ superconductors, it is necessary to introduce massless Dirac fermions, which results from the $d-wave$ pairing symmetry.
An essential point is that such massless Dirac fermions have statistical interactions with vortices, described by an effective $BF$ term with a statistical angle $\pi$, which implies mutual semionic statistics \cite{Senthil:1999czm, Franz:2002qy}.
An effective field theory for $d-wave$ superconductors is given by the path integral expression for massless Dirac fermions $\psi_{n \sigma}$, Cooper-pair vortices $\Phi_{cp}$, superfluid fluctuations $b_{\mu}^{cp}$, and emergent $U(1)$ statistical gauge fields $c_{\mu}$,
\bqa && Z_{\rm d-wave}
= \int D \psi_{n \sigma} D \Phi_{cp} D b_{\mu}^{cp} D c_{\mu} \exp\Big( - S_{\text{Dual}}^{\rm d-wave} \Big) , \nn \eqa
where the effective action is
\bqa && S_{\rm Dual}^{\rm d-wave}
= \int_{0}^{\beta} d \tau \int d^{2} x \Big\{ \bar{\psi}_{n \sigma} \gamma_{\mu} (\partial_{\mu} - i c_{\mu}) \psi_{n \sigma} \nn && + \frac{1}{2 g_{c}^{2}} (\epsilon_{\mu\nu\lambda} \partial_{\nu} c_{\lambda})^{2} + \frac{i}{\pi} \epsilon_{\mu\nu\lambda} b_{\mu}^{cp} \partial_{\nu} c_{\lambda} + |(\partial_{\mu} - i b_{\mu}^{cp}) \Phi_{cp}|^{2} \nn && + m_{cp}^{2} |\Phi_{cp}|^{2} + \frac{u_{cp}}{2} |\Phi_{cp}|^{4} + \frac{1}{2 \rho_{cp}} (\epsilon_{\mu\nu\lambda} \partial_{\nu} b_{\lambda}^{cp})^{2} \Big\} . \eqa
Here, $n$ and $\sigma$ in $\psi_{n \sigma}$ represent spin and flavor quantum numbers, respectively, where the flavor number is determined by the $d-wave$ pairing symmetry, for example, $n = 2$ in a square lattice. The $BF$ term between $b_{\mu}^{cp}$ and $c_{\mu}$ describes the mutual semionic statistics between $\Phi_{cp}$ and $\psi_{n \sigma}$ with $\theta = \pi$ in $\frac{i}{\theta} \epsilon_{\mu\nu\lambda} b_{\mu}^{cp} \partial_{\nu} c_{\lambda}$.

To investigate the topological nature of the superconducting phase, we consider the case when vortices are gapped.
Then, we obtain the following effective field theory in the $d-wave$ superconducting phase
\bqa && Z_{\rm dSC} = \int D \psi_{n \sigma} D b_{\mu}^{cp} D c_{\mu} \nn && \exp\Big[ - \int_{0}^{\beta} d \tau \int d^{2} x \Big\{ \bar{\psi}_{n \sigma} \gamma_{\mu} (\partial_{\mu} - i c_{\mu}) \psi_{n \sigma} \nn && + \frac{1}{2 \rho_{cp}} (\epsilon_{\mu\nu\lambda} \partial_{\nu} b_{\lambda}^{cp})^{2} + \frac{i}{\pi} \epsilon_{\mu\nu\lambda} b_{\mu}^{cp} \partial_{\nu} c_{\lambda} + \frac{1}{2 g_{c}^{2}} (\epsilon_{\mu\nu\lambda} \partial_{\nu} c_{\lambda})^{2} \Big\} \Big] , \nn \eqa
where the phase stiffness parameter $\rho_{cp}$ acquires some renormalization from gapped vortex excitations.
Compared with the canonical expression discussed in the previous section
\bqa && S_{\rm Dual}^{\rm dSC} = \int d^{3} x \Big\{ \frac{1}{8\pi} [\bm{H}^{-1}]_{ff'} (\epsilon_{\mu\nu\lambda} \partial_{\nu} b_{\lambda}^{f}) (\epsilon_{\mu\nu'\lambda'} \partial_{\nu'} b_{\lambda'}^{f'}) \nn && + \frac{K_{fc}}{2\pi} \epsilon_{\mu\nu\lambda} b_{\mu}^{f} \partial_{\nu} a_{\lambda}^{c} + \frac{1}{2} G_{cc'} (\epsilon_{\mu\nu\lambda} \partial_{\nu} a_{\lambda}^{c}) (\epsilon_{\mu\nu'\lambda'} \partial_{\nu'} a_{\lambda'}^{c'}) \nn && + \bar{\psi}_{n \sigma} \gamma_{\mu} (\partial_{\mu} - i a_{\mu}^{c}) \psi_{n \sigma} \Big\} , \nonumber \eqa
we have identification of
\bqa && b_{\mu}^{f} = b_{\mu}^{cp} \delta_{f 1} , ~~~~~ a_{\mu}^{c} = c_{\mu} \delta_{c 1} ,
\eqa
for dual photon and statistical photon fields, respectively,
\bqa && [\bm{H}^{-1}]_{ff'} = \frac{4\pi}{\rho_{cp}} \delta_{f 1} \delta_{f' 1} ,
\eqa
for the phase stiffness parameter,
\bqa && [\bm{K}]_{fc} = 2 \delta_{f 1} \delta_{c 1} \longrightarrow [\bm{K}^{-1}]_{11} = \frac{1}{2} ,
\eqa
for the $\bm{K}$ matrix, and
\bqa && G_{cc'}^{a} = \frac{1}{g_{c}^{2}} \delta_{c 1} \delta_{c' 1},
 \eqa
for the gauge dynamics. Here, the $\bm{K}$ matrix is just a number.
In this respect it seems that there do not exist any massless modes except for fermion degrees of freedom. We will discuss this point below more carefully.

It is straightforward to see that this effective dual action has a discrete one-form symmetry
\bqa
&&
b_{\mu}^{f} \mapsto
b_{\mu}^{f} + q_c^T [\bm{K}^{-1}]_{cf} \lambda_{\mu} , \nn
&&
\partial_{\mu} \lambda_{\nu} - \partial_{\nu} \lambda_{\mu} = 0 ,
~~~~~ \int_{C} d l_{\mu} \lambda_{\mu} \in 2 \pi Z . \nonumber \eqa
The curvature-free condition of the one-form gauge field makes the kinetic-energy term be invariant automatically. The $BF$ term is also invariant in the following way
\bqa \delta_{1} S_{\rm Dual}^{\rm d-wave}
&=&
\frac{1}{2\pi} q^T_c [\bm{K}^{-1} \bm{K}]_{cc'}
\int d^{3} x \epsilon_{\mu\nu\lambda} \lambda_{\mu} \partial_{\nu} a_{\lambda}^{c'} \nn &\in& 2 \pi Z . \nonumber \eqa

One can show that the Wilson line of the dual photon field $V_{p}(C) = \exp\Big( i p_{f} \int_{C} d l_{\mu} b_{\mu}^{f} \Big)$ exhibits the perimeter law, implying that this $Z_{2}$ global one-form symmetry is broken for the ground state. 
The braiding statistics of the Wilson loop and the vortex operator is given by 
\bqa &&
\frac{\Big\langle W_{q}(C) V_{p}(C') \Big\rangle}{\Big\langle W_{q}(C) \Big\rangle \Big\langle V_{p}(C') \Big\rangle} \nn &&
= \exp\Big[ - 2 \pi i (q^T_{c} [\bm{K}^{-1}]_{cf} p_{f}) ~ Lk(C,C') \Big] . \nonumber
\eqa
The factor $q^T_{c} [\bm{K}^{-1}]_{cf} p_{f}$ is multiple of $1/2$, 
and the statistics is semionic. 

However, the above discussion is based on incomplete treatment for massless Dirac fermions, i.e., considering
\bqa && \mathcal{Z} = \int D b_{\mu}^{cp} D c_{\mu} \exp\Big[ - \int_{0}^{\beta} d \tau \int d^{2} x \Big\{ \frac{1}{2 \rho_{cp}} (\epsilon_{\mu\nu\lambda} \partial_{\nu} b_{\lambda}^{cp})^{2} \nn && + \frac{i}{\pi} \epsilon_{\mu\nu\lambda} b_{\mu}^{cp} \partial_{\nu} c_{\lambda} + \frac{1}{2 g_{c}^{2}} (\epsilon_{\mu\nu\lambda} \partial_{\nu} c_{\lambda})^{2} \Big\} \Big] , \eqa
where such Dirac fermions were assumed to be massive through spontaneous ``chiral symmetry breaking" \cite{Appelquist:1986fd, Pisarski:1984dj}, for example, and thus irrelevant at low energies. If we consider the limit of large number of flavors, we have an effective field theory which describes a conformal invariant fixed point in three spacetime dimensions \cite{Hermele:2005dkq}, given by
\bqa && \mathcal{L}_{\rm dSC} = N_{f} \Big\{ \frac{64}{\pi^{2}} (\epsilon_{\mu\nu\lambda} \partial_{\nu} b_{\lambda}^{cp}) \frac{1}{\sqrt{- \partial^{2}}} (\epsilon_{\mu\nu'\lambda'} \partial_{\nu'} b_{\lambda'}^{cp}) \nn && + \frac{i}{\pi} \epsilon_{\mu\nu\lambda} b_{\mu}^{cp} \partial_{\nu} c_{\lambda} + \frac{1}{16} (\epsilon_{\mu\nu\lambda} \partial_{\nu} c_{\lambda}) \frac{1}{\sqrt{- \partial^{2}}} (\epsilon_{\mu\nu'\lambda'} \partial_{\nu'} c_{\lambda'}) \Big\} . \nn \label{LFL_DSC_TO} \eqa
The nonlocal expression for the kinetic-energy term of gauge fields should be understood in the momentum space.
Here, the Maxwell dynamics for both kinetic energies of gauge fields are irrelevant and neglected in this expression.
In addition, $b_{\lambda}^{cp}$ has been scaled as $N_{f} b_{\lambda}^{cp}$.
We emphasize that this is a self-consistent effective Lagrangian in the large-flavor limit, which may be regarded to be classical.
Although the existence of massless fermions break the discrete one-form symmetry of the theory without fermions,
one can see that $Z_{2}$ one-form global symmetry emerges in this limit.
Based on this critical field theory, one may calculate both Wilson's lines and find the perimeter law, which results from the deconfined nature of both quasiparticles and vortices involved with their long-range effective interactions.
In other words, the $Z_{2}$ one-form global symmetry is broken spontaneously in the presence of massless Dirac fermions at least in the large$-N_{f}$ limit.
Based on this observation, we claim that the system acquires $Z_2$ topological order in this phase.

\section{
Topological property of $d-wave$ superconductors from doped Mott insulators
}\label{sec:d-wave-mott}

\subsection{High $T_{c}$ cuprates as doped Mott insulators}

\subsubsection{Effective UV lattice Hamiltonian}

One way to see high $T_{c}$ cuprate superconductors is
that they are from doped Mott insulators \cite{Zhang:1988yua}.
To discuss this aspect, we introduce an effective lattice Hamiltonian, which consists of $Cu$ and $O$ effective lattice Hamiltonians
\bqa && H_{UV} = H_{Cu} + H_{O} + H_{Cu-O} . \eqa
Here, the local Hamiltonian for the $Cu$ site is given by
\bqa && H_{Cu} = \epsilon_{d} \sum_{i = (i_{x}, i_{y})} d_{i\sigma}^{\dagger} d_{i\sigma} + U \sum_{i = (i_{x}, i_{y})} n_{i\uparrow}^{d} n_{i\downarrow}^{d} . \eqa
$d_{i\sigma}$ is an electron annihilation operator at the $Cu$ site $i$, where it acts on the $d_{x^{2}-y^{2}}$ orbital for $Cu^{2+}$ with its local energy $\epsilon_{d}$. $U$ is an effective Coulomb interaction energy for the $d_{x^{2} - y^{2}}$ orbital of $Cu^{2+}$. $n_{i\sigma}^{d} = d_{i\sigma}^{\dagger} d_{i\sigma}$ is an electron density operator of spin $\sigma$. $i = (i_{x}, i_{y})$ denotes a square lattice.
The local Hamiltonian for the $O$ site is given by
\bqa && H_{O} = \epsilon_{p} \sum_{i = (i_{x}, i_{y})} \Big( p_{i+x/2 \sigma}^{x \dagger} p_{i+x/2 \sigma}^{x} + p_{i+y/2 \sigma}^{y \dagger} p_{i+y/2 \sigma}^{y} \Big) . \nn \eqa
$p_{i+x/2 \sigma}^{x}$ ($p_{i+y/2 \sigma}^{y}$) is an electron annihilation operator at the $O$ site $i+x/2$ ($i+y/2$), where it acts on the $p_{x}$ ($p_{y}$) orbital for $O^{2-}$ with its local energy $\epsilon_{p}$. $O$ sites are in the middle of $Cu$ sites on the square lattice, where $|x| = |y| = 1$ with $(x,y)$. Hopping of electrons is realized by the hybridization between the $Cu$ $d_{x^{2} - y^{2}}$ orbital and the $O$ $p_{x}$ and $p_{y}$ orbitals, given by
\bqa && H_{Cu-O} = - t_{pd} \sum_{i = (i_{x}, i_{y})} \Big( d_{i\sigma}^{\dagger} p_{i+x/2 \sigma}^{x} + H.c. \nn && + d_{i\sigma}^{\dagger} p_{i+y/2 \sigma}^{y} + H.c. \Big) - t_{pd} \sum_{i = (i_{x}, i_{y})} \Big( d_{i\sigma}^{\dagger} p_{i-x/2 \sigma}^{x} + H.c. \nn && + d_{i\sigma}^{\dagger} p_{i-y/2 \sigma}^{y} + H.c. \Big) . \eqa
Here, $t_{pd}$ is strength of the $dp$ hybridization, referred to as an overlap integral of wave functions.

Based on this effective lattice Hamiltonian, one may consider
\bqa && U/t_{pd} \rightarrow \infty . \eqa
This limiting procedure gives rise to the following effective Hamiltonian \cite{Zhang:1988yua}, referred to as the $t-J$ model
\bqa && H_{eff} = - t \sum_{ij} (\tilde{c}_{i\sigma}^{\dagger} \tilde{c}_{j\sigma} + H.c.) + J \sum_{ij} \Big( \bm{S}_{i} \cdot \bm{S}_{j} - \frac{1}{4} n_{i} n_{j} \Big) . \nn \eqa
Here, $\tilde{c}_{i\sigma} = P_{G} c_{i\sigma} P_{G}$ is an electron annihilation operator, where $P_{G} = \Pi_{j} (1 - n_{i\uparrow} n_{i\downarrow})$ is the Gutzwiller projection operator to extract out double occupancy sites. At half filling, the kinetic energy term vanishes identically. In other words, hopping of these electrons can be realized only when inter sites are empty. More precisely, $\tilde{c}_{i\sigma}$ represents a doped hole, referred to as a Zhang-Rice singlet \cite{Zhang:1988yua}. Since $\epsilon_{p} > \epsilon_{d}$ is satisfied for high $T_{c}$ cuprates, an electron can be extracted out from the $O$ site via hole doping. Such a doped hole turns out to form a singlet pair with the hole of the $Cu$ site for the $d^{9}$ configuration. Although the term of the Zhang-Rice singlet explains the UV origin precisely, we just use the term of electrons for convenience. These electrons have superexchange interactions, given by the last Heisenberg term. $\bm{S}_{i} = \frac{1}{2} c_{i\alpha}^{\dagger} \bm{\sigma}_{\alpha\beta} c_{i\beta}$ is a spin operator and $n_{i} = n_{i\uparrow} + n_{i\downarrow}$ is a density operator with $n_{i\sigma} = c_{i\sigma}^{\dagger} c_{i\sigma}$. The Gutzwiller projection operator guarantees the following constraint
\bqa && n_{i} \leq 1 , \label{Double_Occupancy_Constraint} \eqa
which implies that the low-energy effective Hamiltonian is to describe physics of doped Mott insulators.

\subsubsection{U(1) slave-boson representation and effective U(1) lattice gauge theory}

To solve the double occupancy constraint given by the Gutzwiller projection operator, U(1) slave-boson representation has been introduced
\bqa && c_{i\sigma} = b_{i}^{\dagger} f_{i\sigma} . \label{U1SB_representation} \eqa
Here, the electron annihilation operator is represented by a composite operator, given by a holon field $b_{i}^{\dagger}$ to carry an electron charge quantum number and a spinon field $f_{i\sigma}$ to carry an electron spin quantum number. Then, the constraint equation (\ref{Double_Occupancy_Constraint}) is expressed as
\bqa && b_{i}^{\dagger} b_{i} + f_{i\sigma}^{\dagger} f_{i\sigma} = 1 , \label{Single_Occupancy_Constraint} \eqa
which allows either one holon or one spinon with spin $\sigma$ at the site $i$.

Now, it is straightforward to construct the path integral expression for the partition function
\bqa && Z = \int D f_{i\sigma} D b_{i} D \lambda_{i} \exp\Big[ - \int_{0}^{\beta} d \tau \Big\{ \sum_{i} f_{i\sigma}^{\dagger} (\partial_{\tau} - \mu \nn && - i \lambda_{i}) f_{i\sigma} + \sum_{i} b_{i}^{\dagger} (\partial_{\tau} - i \lambda_{i}) b_{i} + i \sum_{i} \lambda_{i} \nn && - t \sum_{ij} (f_{i\sigma}^{\dagger} b_{i} b_{j}^{\dagger} f_{j\sigma} + H.c.) + J \sum_{ij} \Big( \bm{S}_{i} \cdot \bm{S}_{j} - \frac{1}{4} n_{i} n_{j} \Big) \Big\} \Big] . \nn \label{EFT_U1SB} \eqa
Here, $\lambda_{i}$ is a Lagrange multiplier field to impose the single occupancy constraint Eq. (\ref{Single_Occupancy_Constraint}). $\mu$ is an electron chemical potential to determine hole concentration given by
$\delta = \langle b_{i}^{\dagger} b_{i} \rangle$.
The density and spin operators are given by
\bqa && n_{i\sigma} = f_{i\sigma}^{\dagger} f_{i\sigma} , ~~~~~ \bm{S}_{i} = \frac{1}{2} f_{i\alpha}^{\dagger} \bm{\sigma}_{\alpha\beta} f_{i\beta} , \label{Density_Spin_U1SB} \eqa
respectively, where the constraint equation has been utilized.

Introducing Eq. (\ref{Density_Spin_U1SB}) into Eq. (\ref{EFT_U1SB}) and performing the Hubbard-Stratonovich decomposition for interactions into spin singlet channels, we find an effective field theory
\bqa && Z = \int D f_{i\sigma} D b_{i} D \lambda_{i} D \chi_{ij}^{b} D \chi_{ij}^{f} D \Delta_{ij} D \varphi_{i} \nn && \exp\Big[ - \int_{0}^{\beta} d \tau \Big\{ \sum_{ij} f_{i\sigma}^{\dagger} \Big( (\partial_{\tau} - \mu - i \lambda_{i}) \delta_{ij} - \frac{J}{2} \varphi_{j} \Big) f_{i\sigma} \nn && - t \sum_{ij} f_{i\sigma}^{\dagger} \chi_{ij}^{b} f_{j\sigma} - H.c. - \frac{J}{4} \sum_{ij} \Delta_{ij}^{\dagger} (\epsilon_{\alpha\beta} f_{i\alpha} f_{j\beta}) - H.c. \nn && + \sum_{i} b_{i}^{\dagger} (\partial_{\tau} - i \lambda_{i}) b_{i} - t \sum_{ij} b_{i} \chi_{ij}^{f} b_{j}^{\dagger} - H.c. + i \sum_{i} \lambda_{i} \nn && + t \sum_{ij} \chi_{ij}^{b} \chi_{ij}^{f} + H.c. + \frac{J}{4} \sum_{ij} \Big( - |\chi_{ij}^{f}|^{2} + |\Delta_{ij}|^{2} \Big) \nn && + \frac{J}{4} \sum_{ij} \varphi_{i} \varphi_{j} \Big\} \Big] . \eqa
$\chi_{ij}^{b}$ and $\chi_{ij}^{f}$ are inter-site particle-hole composite fields to give effective bands to holons and spinons, respectively. $\Delta_{ij}$ is a link variable to describe $d-wave$ pairing of spinon Cooper pairs, which is our central object. $\varphi_{i}$ is an effective potential to decompose density-density interactions. We refer one way of this construction to Ref. \cite{Lee:2006zzc}.

To discuss possible low energy physics of this effective theory, we focus on transverse fluctuations for these ``order parameters" given by
\bqa && \chi_{ij}^{b} = \chi_{ij}^{b s} e^{i a_{ij}} , ~~~~~ \chi_{ij}^{f} = {\chi}_{ij}^{f s} e^{- i a_{ij}} , ~~~~~ \Delta_{ij} = \Delta_{ij}^{s} e^{i \phi_{ij}} . \nn \eqa
Here, $a_{ij}$ is a phase field for the hopping order parameter, which plays the role of $U(1)$ gauge fields.
These phase fluctuations contribute to lowering the ground state energy by spontaneous generation of such $U(1)$ internal magnetic fluxes in the lattice.
This $U(1)$ gauge redundancy may be regarded to originate from the $U(1)$ slave-boson representation itself, where the electron annihilation operator of Eq. (\ref{U1SB_representation}) is invariant under $U(1)$ gauge transformation of $f_{i\sigma} \rightarrow e^{i \theta_{i}} f_{i\sigma}$ and $b_{i} \rightarrow e^{i \theta_{i}} b_{i}$. $\phi_{ij}$ is a phase field for the $d-wave$ spinon pairing order parameter, which carries the internal $U(1)$ gauge charge $2 e$, i.e., transformed as $\phi_{ij} \rightarrow \phi_{ij} + \theta_{i} + \theta_{j}$ under the $U(1)$ gauge transformation.

Considering
\bqa && i \lambda_{i} = \lambda_{i}^{s} + i a_{i \tau} , ~~~~~ \varphi_{i} = \varphi_{i}^{s} + \delta \varphi_{i} , \eqa
where the superscript $s$ means a saddle-point value, and introducing
\bqa && - \frac{J}{2} \sum_{j \in i} \varphi_{j}^{s} - i a_{i \tau} - \frac{J}{2} \sum_{j \in i} \delta \varphi_{j} \rightarrow - i a_{i \tau} \eqa
into the resulting lattice model, we construct an effective lattice gauge theory as follows
\bqa && Z = \int D f_{i\sigma} D b_{i} D a_{i \tau} D a_{ij} D \phi_{ij} \nn && \exp\Big[ - \int_{0}^{\beta} d \tau \Big\{ \sum_{i} f_{i\sigma}^{\dagger} \Big( \partial_{\tau} - \mu - \lambda_{i}^{s} - i a_{i \tau} \Big) f_{i\sigma} \nn && - t \sum_{ij} \chi_{ij}^{b s} f_{i\sigma}^{\dagger} e^{i a_{ij}} f_{j\sigma} - H.c. \nn && - \frac{J}{4} \sum_{ij} \Delta_{ij}^{s \dagger} e^{- i \phi_{ji}} (\epsilon_{\alpha\beta} f_{i\alpha} f_{j\beta}) - H.c. \nn && + \sum_{i} b_{i}^{\dagger} (\partial_{\tau} - \lambda_{i}^{s} - i a_{i \tau}) b_{i} - t \sum_{ij} \chi_{ji}^{f s \dagger} b_{j}^{\dagger} e^{i a_{ji}} b_{i} - H.c. \nn && + i \sum_{i} a_{i \tau} + \sum_{i} \lambda_{i}^{s} + t \sum_{ij} \chi_{ij}^{b s} \chi_{ij}^{f s} + H.c. \nn && + \frac{J}{4} \sum_{ij} \Big( - |\chi_{ij}^{f s}|^{2} + |\Delta_{ij}^{s}|^{2} \Big) \Big\} \Big] . \eqa
Here, $a_{i\tau}$ is a time-component of the $U(1)$ gauge field.
Based on this $U(1)$ lattice gauge theory, we construct an effective field theory below. We refer this procedure to Ref. \cite{Lee:2006zzc}.

\subsection{Absence of topological order in $s-wave$ superconductors from doped Mott insulators}

First, we consider the case of $s-wave$ superconductors from doped Mott insulators, where spinons are gapped to be neglected at low energies. Then, the corresponding effective field theory is given by
\bqa && S_{\rm MI}^{\rm s-wave}
= \int_{0}^{\beta} d \tau \int d^{2} x \Big\{ \frac{\rho_{sp}}{2} | \partial_{\mu} \phi_{sp} - 2 a_{\mu} |^{2} \nn && + \frac{\rho_{b}}{2} | \partial_{\mu} \phi_{b} - a_{\mu} |^{2} + \frac{1}{2 g^{2}} (\epsilon_{\mu\nu\lambda} \partial_{\nu} a_{\lambda})^{2} \Big\} . \eqa
$\rho_{sp}$ ($\rho_{b}$) is the phase stiffness parameter for the dynamics of spinon Cooper pairs (holons), given by $\rho_{sp} \sim | \langle \Delta_{sp} \rangle |^{2}$ ($\rho_{b} \sim | \langle b \rangle |^{2}$) with $\Delta_{sp} = \langle \Delta_{sp} \rangle e^{i \phi_{sp}}$ ($b = \langle b \rangle e^{i \phi_{b}}$).
$g$ is the unit of the internal $U(1)$ gauge charge. A noticeable point is that the $U(1)$ gauge charge for spinon Cooper pairs is $2 g$ while that of holons is $g$.

Performing the duality transformation, we obtain an effective dual field theory
\bqa && Z_{\rm MI}^{\rm s-wave}
= \int D \Phi_{b} D \Phi_{sp} D a_{\mu} D b_{\mu}^{sp} D b_{\mu}^{b} \nn && \exp\Big[ - \int_{0}^{\beta} d \tau \int d^{2} x \Big\{ |(\partial_{\mu} - i b_{\mu}^{sp}) \Phi_{sp}|^{2} \nn && + m_{sp}^{2} |\Phi_{sp}|^{2} + \frac{u_{sp}}{2} |\Phi_{sp}|^{4} \nn && + \frac{1}{2 \rho_{sp}} (\epsilon_{\mu\nu\lambda} \partial_{\nu} b_{\lambda}^{sp})^{2} + \frac{i}{\pi} \epsilon_{\mu\nu\lambda} b_{\mu}^{sp} \partial_{\nu} a_{\lambda} + \frac{1}{2 g^{2}} (\epsilon_{\mu\nu\lambda} \partial_{\nu} a_{\lambda})^{2} \nn && + |(\partial_{\mu} - i b_{\mu}^{b}) \Phi_{b}|^{2} + m_{b}^{2} |\Phi_{b}|^{2} + \frac{u_{b}}{2} |\Phi_{b}|^{4} \nn && + \frac{1}{2 \rho_{b}} (\epsilon_{\mu\nu\lambda} \partial_{\nu} b_{\lambda}^{b})^{2} + \frac{i}{2\pi} \epsilon_{\mu\nu\lambda} b_{\mu}^{b} \partial_{\nu} a_{\lambda} \Big\} \Big] . \label{SSC_MI_Dual_EFT} \eqa
$\Phi_{sp}$ ($\Phi_{b}$) is a vortex field for the spinon Cooper pair (holon) field and $b_{\mu}^{sp}$ ($b_{\mu}^{b}$) is the corresponding superfluid sound mode. Such spinon pair and holon sound modes are coupled to internal $U(1)$ gauge fields, described by $BF$ terms with appropriate $U(1)$ charges.

Here, we focus on the superconducting phase, where both spinon pair and holon vortices are gapped. Then, the effective dual field theory is given by
\begin{widetext}
\bqa && Z_{\rm MI}^{\rm s-SC} = \int D a_{\mu} D b_{\mu}^{sp} D b_{\mu}^{b} \exp\Big[ - \int_{0}^{\beta} d \tau \int d^{2} x \Big\{ \frac{1}{2 \rho_{sp}} (\epsilon_{\mu\nu\lambda} \partial_{\nu} b_{\lambda}^{sp})^{2} + \frac{1}{2 \rho_{b}} (\epsilon_{\mu\nu\lambda} \partial_{\nu} b_{\lambda}^{b})^{2} \nn && + \frac{i}{\pi} \epsilon_{\mu\nu\lambda} b_{\mu}^{sp} \partial_{\nu} a_{\lambda} + \frac{i}{2\pi} \epsilon_{\mu\nu\lambda} b_{\mu}^{b} \partial_{\nu} a_{\lambda} + \frac{1}{2 g^{2}} (\epsilon_{\mu\nu\lambda} \partial_{\nu} a_{\lambda})^{2} \Big\} \Big] . \label{SSC_from_MI} \eqa
\end{widetext}
Compared with Eq. (\ref{Dual_Field_Theory_CFL}), we obtain
\bqa
\bm{H} = \begin{pmatrix} \frac{\rho_{sp}}{4 \pi} & 0 \\ 0 & \frac{\rho_{b}}{4 \pi} \end{pmatrix}
\eqa
for the stiffness parameter,
\bqa
\bm{K} = \begin{pmatrix} 2 \\ 1 \end{pmatrix}
\eqa
for the $\bm{K}$ matrix, and
\bqa && G_{cc'} = \frac{1}{g^{2}} \delta_{c 1} \delta_{c' 1} \eqa
for the gauge coupling constant.

Since the number of superfluid sound modes is larger than that of U(1) gauge fields, it is natural to expect the existence of neutral superfluid modes, which decouple from U(1) gauge fields. Indeed, it exists, given by
\bqa && [\bm{b}_{0}]_{\mu}^{f} = P_{ff'}^{b} [\bm{b}]_{\mu}^{f'} , \nonumber \eqa
where the projection operator is
\bqa && P_{ff'}^{b} = \delta_{ff'} - [\bm{K} \bm{K}^{+}]_{ff'} \nonumber \eqa
with the Moore-Penrose inverse of $\bm{K}$
\bqa \bm{K}^{+} = \begin{pmatrix} \frac{2}{5} & \frac{1}{5} \end{pmatrix} . \eqa
More explicitly, such a neutral superfluid sound mode is given by
\bqa && [\bm{b}_{0}]_{\mu}^{f} = \frac{1}{5} (b_{\mu}^{sp} - 2 b_{\mu}^{b}) \delta_{f 1} . \eqa

Now, let us study the one-form global symmetry of Eq. (\ref{SSC_from_MI}).
The dual effective field theory of the $s-wave$ superconducting phase from a spin liquid state has the following one-form global symmetry
\bqa
 && b_{\mu}^{f}
 \mapsto
 b_{\mu}^{f} + [\bm{K}^{+}]_{cf} \lambda_{\mu}^{c} , \nonumber \eqa
where $\lambda_{\mu}^{c}$ is a flat connection satisfying $\partial_{\mu} \lambda_{\nu}^{c} - \partial_{\nu} \lambda_{\mu}^{c} = 0$
with normalization $\int_{C} d l_{\mu} \lambda_{\mu}^{c} \in 2 \pi Z$.
From the explicit form of the Moore-Penrose inverse of $\bm{K}$,
one can see that it is a $Z_{5}$ one-form symmetry.

This state also has a continuous one-form symmetry,
\bqa &&
b_{\mu}^{f}
\mapsto
b_{\mu}^{f}
+ \epsilon [\bm{D}^{\bar{\alpha}}]_{f} \rho_{\mu} ,
\label{eq:u1-one-form}
\eqa
where $\epsilon$ is a continuous parameter,
$\rho_{\mu}$ is a flat one-form connection
normalized as $\int_{C} d l_{\mu} \rho_{\mu} \in 2 \pi Z$,
and
$\bm{D}^{\bar{\alpha}}$ is the element of the cokernel of $\bm{K}$,
\bqa
&& [\bm{D}^{\bar{\alpha}}]_{f} = \begin{pmatrix} - 1 & 2 \end{pmatrix} .
\eqa
Here, the cokernel is one-dimensional.
Under the transformation (\ref{eq:u1-one-form}),
Wilson loop operators are transformed as
\bqa && V_{p}(C)
\mapsto
\exp\Big( 2 \pi i p_{f} \, \epsilon [\bm{D}^{\bar{\alpha}}]_{f} \Big) V_{p}(C) .
\eqa

There is no discrete one-form symmetry to shift the gauge field $a$.
Recall that a generic form of discrete one-form transformation takes the form
\bqa
 && a_{\mu}^{c} \mapsto
 a_{\mu}^{c} + [\bm{K}^{+}]_{cf} p_f \lambda_{\mu},
 \label{eq:one-form-a}
 \eqa
where $p_f \in (\cok \, \bm K )^\perp$.
The charge vector $p_f$ should be chosen to be orthogonal to $\bm D^{\bar\alpha}$
so that the generator is a topological operator.
One choice with minimal integer is
\be
\bm p =
\begin{pmatrix}
2 & 1
\end{pmatrix} .
\ee
Although the transformation (\ref{eq:one-form-a}) with this charge vector
does not change the action (up to $2\pi Z$), this is not a symmetry of the system,
because this acts trivially on the Wilson loop of $a$.

In the current case,
the $Z_{5}$ one-form symmetry is a subgroup of the $U(1)$ continuous one-form symmetry (\ref{eq:u1-one-form}).
Under the $Z_5$ one-form symmetry,
the Wilson loop operator for the spinon-pair gauge field is transformed as
\bqa && V_{sp} = \exp\Big( i  \int_{C} d l_{\mu} b_{\mu}^{sp} \Big) \nn && \mapsto \exp\Big( \frac{2}{5} \times 2 \pi i \Big) V_{sp} \eqa
while that of the holon gauge field is changed as
\bqa && V_{b} = \exp\Big( i \int_{C} d l_{\mu} b_{\mu}^{b} \Big) \nn && \mapsto \exp\Big( \frac{1}{5} \times 2 \pi i \Big) V_{b} = \exp\Big( - \frac{4}{5} \times 2 \pi i \Big) V_{b} .
\eqa
Thus, this symmetry acts as $Z_{5}$ phase rotation on those operators.
In fact, this phase rotation can also be generated as a $U(1)$ one-form symmetry
(\ref{eq:u1-one-form}),
\bqa && b_{\mu}^{sp} \mapsto b_{\mu}^{sp} + \epsilon (- 1) \rho_{\mu} , ~~~~~ b_{\mu}^{b} \mapsto b_{\mu}^{b} + \epsilon (2) \rho_{\mu}
\eqa
by choosing
\bqa && \epsilon = - \frac{2}{5} . \eqa
Hence, the $Z_{5}$ one-form symmetry is a subgroup of the $U(1)$ continuous one-form symmetry, if we look at the transformation property of physical operators.
In three spacetime dimensions, a continuous one-form symmetry cannot be broken, guaranteed by the generalized Coleman-Mermin-Wagner theorem for higher-form symmetries \cite{gaiotto2015generalized}.
Therefore, the $Z_{5}$ one-form symmetry cannot be broken either.
As a result, the $s-wave$ superconducting phase with one neutral superfluid sound mode, resulting from a doped Mott insulating state, does not have a topological order.
This implies that conventional $s-wave$ superconductivity, induced by the condensation of electron Cooper pairs (identified as superfluidity here),
can be smoothly connected with the $s-wave$ superconducting phase with one neutral superfluid sound mode, resulting from the spin liquid state.

Even though there is no topological order in this superconducting state,
the mutual statistics between vortices and quasiparticles can be exotic.
Let us denote $W(C)\equiv \exp \left( i \int_C d l_\mu a_\mu \right)$.
The correlation functions of $W(C)$ and $V_{sp}(C)$, $V_{b}(C)$ satisfy
\be
\langle
W(C) V_{sp}(C')
\rangle
= e^{- \frac{4 \pi i }5  Lk(C, C')}
\langle
V_{sp}(C')
\rangle ,
\label{eq:linking1}
\ee
\be
\langle
W(C) V_{b}(C')
\rangle
= e^{- \frac{2\pi i }5 Lk (C, C')}
\langle
V_{b}(C')
\rangle ,
\label{eq:linking2}
\ee
respectively, where we used the fact that Wilson loops associated with $a_\mu$ are topological, and
\be
\langle W_q (C) \rangle = 1 .
\ee
The relations (\ref{eq:linking1}, \ref{eq:linking2})
indicate that the Wilson loop $W_q(C)$ is the generator of the
$Z_5$ one-form symmetry and
$V_p(C)$ is the corresponding charged object.
Although the $Z_5$ one-form symmetry is not spontaneously broken,
this braiding phase is an observable effect.

\subsection{
$d-wave$ superconductors from doped Mott insulators at large$-N_{f}$
}

Finally, we discuss the topological properties of $d-wave$ superconductors from doped Mott insulators, based on the perspectives of higher-form symmetries.
Introducing massless Dirac fermions into the effective dual field theory Eq. (\ref{SSC_MI_Dual_EFT}) for $s-wave$ superconductors from doped Mott insulators, we obtain
\begin{widetext}
\bqa && Z_{\rm MI}^{\rm d-wave}
=
\int D \psi_{n \sigma} D \Phi_{b} D \Phi_{sp} D a_{\mu} D b_{\mu}^{sp} D b_{\mu}^{b} D c_{\mu} \exp\Big[ - \int_{0}^{\beta} d \tau \int d^{2} x \Big\{ \bar{\psi}_{n \sigma} \gamma_{\mu} (\partial_{\mu} - i c_{\mu}) \psi_{n \sigma} + \frac{i}{\pi} \epsilon_{\mu\nu\lambda} b_{\mu}^{sp} \partial_{\nu} c_{\lambda} \nn && + |(\partial_{\mu} - i b_{\mu}^{sp}) \Phi_{sp}|^{2} + m_{sp}^{2} |\Phi_{sp}|^{2} + \frac{u_{sp}}{2} |\Phi_{sp}|^{4} + \frac{1}{2 \rho_{sp}} (\epsilon_{\mu\nu\lambda} \partial_{\nu} b_{\lambda}^{sp})^{2} + \frac{i}{\pi} \epsilon_{\mu\nu\lambda} b_{\mu}^{sp} \partial_{\nu} a_{\lambda} + \frac{1}{2 g^{2}} (\epsilon_{\mu\nu\lambda} \partial_{\nu} a_{\lambda})^{2} \nn && + |(\partial_{\mu} - i b_{\mu}^{b}) \Phi_{b}|^{2} + m_{b}^{2} |\Phi_{b}|^{2} + \frac{u_{b}}{2} |\Phi_{b}|^{4} + \frac{1}{2 \rho_{b}} (\epsilon_{\mu\nu\lambda} \partial_{\nu} b_{\lambda}^{b})^{2} + \frac{i}{2\pi} \epsilon_{\mu\nu\lambda} b_{\mu}^{b} \partial_{\nu} a_{\lambda} \Big\} \Big] . \eqa
\end{widetext}
Here, $\psi_{n \sigma}$ represents a massless Dirac fermion field resulting from the $d-wave$ pairing symmetry of spinon Cooper pairs. Such massless fermions interact with spinon pair vortices statistically, described by the $BF$ term with the statistical angle $\pi$. In other words, when the Dirac fermion turns around the spinon pair vortex, it acquires the Aharonov-Bohm phase of $\pi$. This dual effective field theory is reduced into
\begin{widetext}
\bqa && Z_{\rm MI}^{\rm dSC}
= \int D \psi_{n \sigma} D a_{\mu} D b_{\mu}^{sp} D b_{\mu}^{b} D c_{\mu} \exp\Big[ - \int_{0}^{\beta} d \tau \int d^{2} x \Big\{ \bar{\psi}_{n \sigma} \gamma_{\mu} (\partial_{\mu} - i c_{\mu}) \psi_{n \sigma} \nn && + \frac{1}{2 \rho_{sp}} (\epsilon_{\mu\nu\lambda} \partial_{\nu} b_{\lambda}^{sp})^{2} + \frac{1}{2 \rho_{b}} (\epsilon_{\mu\nu\lambda} \partial_{\nu} b_{\lambda}^{b})^{2} + \frac{i}{\pi} \epsilon_{\mu\nu\lambda} b_{\mu}^{sp} \partial_{\nu} a_{\lambda} + \frac{i}{2\pi} \epsilon_{\mu\nu\lambda} b_{\mu}^{b} \partial_{\nu} a_{\lambda} + \frac{i}{\pi} \epsilon_{\mu\nu\lambda} b_{\mu}^{sp} \partial_{\nu} c_{\lambda} \nn && + \frac{1}{2 g_{a}^{2}} (\epsilon_{\mu\nu\lambda} \partial_{\nu} a_{\lambda})^{2} + \frac{1}{2 g_{c}^{2}} (\epsilon_{\mu\nu\lambda} \partial_{\nu} c_{\lambda})^{2} \Big\} \Big] \label{dSC_MI_Dual_Higgs} \eqa
\end{widetext}
in the superconducting phase, where both spinon pair and holon vortices are gapped.

Before we study the topological property of Eq. (\ref{dSC_MI_Dual_Higgs}),
we consider the case that Dirac fermions are gapped,
and thus neglected at low energies, described by
\begin{widetext}
\bqa && S_{\rm Dual-MI}^{\rm dSC-CSB}
=
\int d^{3} x \Big\{ \frac{1}{8\pi} [\bm{H}^{-1}]_{ff'} (\epsilon_{\mu\nu\lambda} \partial_{\nu} b_{\lambda}^{f}) (\epsilon_{\mu\nu'\lambda'} \partial_{\nu'} b_{\lambda'}^{f'}) + \frac{K_{fc}}{2\pi} \epsilon_{\mu\nu\lambda} b_{\mu}^{f} \partial_{\nu} a_{\lambda}^{c} + \frac{1}{2} G_{cc'} (\epsilon_{\mu\nu\lambda} \partial_{\nu} a_{\lambda}^{c}) (\epsilon_{\mu\nu'\lambda'} \partial_{\nu'} a_{\lambda'}^{c'}) \Big\} , \nonumber
\eqa
\end{widetext}
where we have correspondences
\bqa && b_{\mu}^{1} = b_{\mu}^{sp} , ~~~~~ b_{\mu}^{2} = b_{\mu}^{b} , \nn && a_{\mu}^{1} = a_{\mu} , ~~~~~ a_{\mu}^{2} = c_{\mu} \eqa
for fields,
\bqa
\bm{H} = \begin{pmatrix} \frac{\rho_{sp}}{4 \pi} & 0 \\ 0 & \frac{\rho_{b}}{4 \pi} \end{pmatrix}
\eqa
for the stiffness matrix,
\bqa \bm{K} = \begin{pmatrix} 2 & 2 \\ 1 & 0 \end{pmatrix} \rightarrow \bm{K}^{-1} = \begin{pmatrix} 0 & 1 \\ 1/2 & -1 \end{pmatrix} \eqa
for the $\bm{K}$ matrix, and
\bqa
\bm{G} = \begin{pmatrix} \frac{1}{g_{a}^{2}} & 0 \\ 0 & \frac{1}{g_{c}^{2}} \end{pmatrix}
\eqa
for the gauge coupling-constant matrix.
Since we have $\di ~ (\ke ~ \bm{K}) = \di ~ ( \cok ~ \bm{K}) = 0$,
there are no massless degrees of freedom in this ``$d-wave$" superconducting phase, where such massless Dirac fermions are gapped due to ``chiral symmetry" breaking \cite{Appelquist:1986fd, Pisarski:1984dj}.
We also note that there is no continuous one-form symmetry in this case because $\bm{K}$ is full-rank.
As a result, this superconducting state has the $Z_{2}$ discrete one-form symmetry. 
Wilson loops obey the perimeter law and this symmetry is spontaneously broken. 
Hence this phase has a $Z_2$ topological order.  
The mutual statistics between spinon-pair vortices and quasiparticles is semionic, which confirms the $Z_{2}$ topological order.

Now, we introduce the role of massless Dirac fermions into the above discussion. If we consider the limit of large number of flavors, we have an effective field theory which describes a conformal invariant fixed point in three spacetime dimensions \cite{Hermele:2005dkq}, given by
\bqa && \mathcal{L}_{\rm Dual-MI}^{\rm dSC}
=
N_{f} \Big\{ \frac{1}{16} (\epsilon_{\mu\nu\lambda} \partial_{\nu} c_{\lambda}) \frac{1}{\sqrt{- \partial^{2}}} (\epsilon_{\mu\nu'\lambda'} \partial_{\nu'} c_{\lambda'}) \nn && + \frac{1}{256} (\epsilon_{\mu\nu\lambda} \partial_{\nu} a_{\lambda}) \frac{1}{\sqrt{- \partial^{2}}} (\epsilon_{\mu\nu'\lambda'} \partial_{\nu'} a_{\lambda'}) \nn && + \frac{64}{\pi^{2}} (\epsilon_{\mu\nu\lambda} \partial_{\nu} b_{\lambda}^{sp}) \frac{1}{\sqrt{- \partial^{2}}} (\epsilon_{\mu\nu'\lambda'} \partial_{\nu'} b_{\lambda'}^{sp}) \nn && + \frac{16}{\pi^{2}} (\epsilon_{\mu\nu\lambda} \partial_{\nu} b_{\lambda}^{b}) \frac{1}{\sqrt{- \partial^{2}}} (\epsilon_{\mu\nu'\lambda'} \partial_{\nu'} b_{\lambda'}^{b}) \nn && + \frac{i}{\pi} \epsilon_{\mu\nu\lambda} b_{\mu}^{sp} \partial_{\nu} a_{\lambda} + \frac{i}{2\pi} \epsilon_{\mu\nu\lambda} b_{\mu}^{b} \partial_{\nu} a_{\lambda} + \frac{i}{\pi} \epsilon_{\mu\nu\lambda} b_{\mu}^{sp} \partial_{\nu} c_{\lambda} \Big\} . \nn \label{U1SL_DSC_TO} \eqa
Here, $b_{\mu}^{sp} \rightarrow N_{f} b_{\mu}^{sp}$ and $b_{\mu}^{b} \rightarrow N_{f} b_{\mu}^{b}$ have been performed.
Again, we point out that this effective field theory is self-consistent and classical in the large number of fermion flavors.
All types of superconducting phase fluctuations and $U(1)$ gauge fields are massless, more precisely, the dynamics of which is conformal invariant.
In spite of this quantum criticality, we point out that all the Wilson loop operators show their perimeter-law behaviors, where both vortices and quasiparticles are deconfined to have effective Coulomb interactions.
For comparison with the absence of massless Dirac fermions, we recall the corresponding effective field theory
\bqa &&
\mathcal{L}_{\rm Dual-MI}^{\rm dSC-CSB} = \frac{1}{2 g_{c}^{2}} (\epsilon_{\mu\nu\lambda} \partial_{\nu} c_{\lambda})^{2}+ \frac{1}{2 g_{a}^{2}} (\epsilon_{\mu\nu\lambda} \partial_{\nu} a_{\lambda})^{2} \nn && + \frac{1}{2 \rho_{sp}} (\epsilon_{\mu\nu\lambda} \partial_{\nu} b_{\lambda}^{sp})^{2} + \frac{1}{2 \rho_{b}} (\epsilon_{\mu\nu\lambda} \partial_{\nu} b_{\lambda}^{b})^{2} \nn && + \frac{i}{\pi} \epsilon_{\mu\nu\lambda} b_{\mu}^{sp} \partial_{\nu} a_{\lambda} + \frac{i}{2\pi} \epsilon_{\mu\nu\lambda} b_{\mu}^{b} \partial_{\nu} a_{\lambda} + \frac{i}{\pi} \epsilon_{\mu\nu\lambda} b_{\mu}^{sp} \partial_{\nu} c_{\lambda} , \nn \nonumber \eqa
where all gauge fields are gapped, and thus $Z_{2}$ topologically ordered. The Maxwell dynamics are replaced with the conformal invariant kinetic energy due to the presence of massless Dirac fermions.
The coexistence of the $Z_{2}$ topological order and quantum criticality is an essential feature for $d-wave$ superconductors in the large number of fermion flavors.

This analysis suggests
that $d-wave$ superconductors from $U(1)$ spin liquids may not be adiabatically connected with those from Landau's Fermi liquids,
in contrast to the case of $s-wave$ superconductors as long as the physics of massless Dirac fermions is governed by the conformal invariant fixed point.
We recall the effective field theory of Eq. (\ref{LFL_DSC_TO}) for $d-wave$ superconductors from Landau's Fermi liquids
\bqa && \mathcal{L}_{\rm dSC} = N_{f} \Big\{ \frac{64}{\pi^{2}} (\epsilon_{\mu\nu\lambda} \partial_{\nu} b_{\lambda}^{cp}) \frac{1}{\sqrt{- \partial^{2}}} (\epsilon_{\mu\nu'\lambda'} \partial_{\nu'} b_{\lambda'}^{cp}) \nn && + \frac{i}{\pi} \epsilon_{\mu\nu\lambda} b_{\mu}^{cp} \partial_{\nu} c_{\lambda} + \frac{1}{16} (\epsilon_{\mu\nu\lambda} \partial_{\nu} c_{\lambda}) \frac{1}{\sqrt{- \partial^{2}}} (\epsilon_{\mu\nu'\lambda'} \partial_{\nu'} c_{\lambda'}) \Big\} . \nonumber \eqa
The difference of the $\bm{K}_{fc}$ matrix between Eq. (\ref{U1SL_DSC_TO}) and Eq. (\ref{LFL_DSC_TO}) indicates that the topological order of Eq. (\ref{U1SL_DSC_TO}) differs from that of Eq. (\ref{LFL_DSC_TO}). We speculate that this topological phase transition is a deconfinement-confinement transition involved with the compact $U(1)$ gauge field $a_{\mu}$, where magnetic monopoles as instantons in three spacetime dimensions play an essential role.

Performing the duality transformation for compact U(1) gauge fields $a_{\mu}$ and taking into account the dilute instanton-gas approximation, we obtain an effective field theory for instanton-type magnetic monopole excitations
\bqa && \mathcal{L}_{\rm Dual-MI}^{\rm dSC}
=
N_{f} \Big\{ \frac{1}{16} (\epsilon_{\mu\nu\lambda} \partial_{\nu} c_{\lambda}) \frac{1}{\sqrt{- \partial^{2}}} (\epsilon_{\mu\nu'\lambda'} \partial_{\nu'} c_{\lambda'}) \nn &&
+ \frac{64}{\pi^{2}} (\epsilon_{\mu\nu\lambda} \partial_{\nu} b_{\lambda}^{sp}) \frac{1}{\sqrt{- \partial^{2}}} (\epsilon_{\mu\nu'\lambda'} \partial_{\nu'} b_{\lambda'}^{sp}) \nn &&
+ \frac{16}{\pi^{2}} (\epsilon_{\mu\nu\lambda} \partial_{\nu} b_{\lambda}^{b}) \frac{1}{\sqrt{- \partial^{2}}} (\epsilon_{\mu\nu'\lambda'} \partial_{\nu'} b_{\lambda'}^{b})
+ \frac{i}{\pi} \epsilon_{\mu\nu\lambda} b_{\mu}^{sp} \partial_{\nu} c_{\lambda} \nn &&
+ \frac{1}{256} (\partial_{\mu} \varphi)
\sqrt{- \p^2 }
(\partial_{\mu} \varphi) - y_{m} \cos \varphi \nn && + \frac{i}{\pi} (\partial_{\mu} \varphi) \Big( b_{\mu}^{sp} + \frac{1}{2} b_{\mu}^{b} \Big) \Big\} . \label{U1SL_DSC_MONOPOLE}
\eqa
Here, $\varphi$ is a scalar magnetic potential field to mediate effective interactions between instanton-type magnetic monopole excitations and $y_{m}$ is an effective fugacity given by $y_m\sim e^{- S_{\rm inst}}$ with a renormalized instanton action $S_{\rm inst}$. Doping dependence would be introduced into the effective instanton action. Although we are not claiming that this procedure is rigorously performed in the presence of massless Dirac fermions, we can argue that the condensation transition of magnetic monopoles may be a Berezinskii-Kosterlitz-Thouless (BKT) type according to one scenario \cite{Kleinert:2002uv, Nogueira:2005aj, Nogueira:2007pn}. If the condensation of such magnetic monopoles occurs, magnetic scalar potential fluctuations are gapped, expected to appear above the optimal doping. Taking into account both magnetic scalar potential fluctuations and ``holon" current fluctuations, we find essentially the same effective field theory as Eq. (\ref{LFL_DSC_TO}), which describes $d-wave$ superconductivity from Landau's Fermi liquids.
This potential existence of a topological phase transition between $d-wave$ superconductors in high $T_{c}$ cuprates, given by a deconfinement-confinement transition, is one of the main implications of the current study and it deserves further studies.

\section{Conclusion}\label{sec:conclusion}

Statistical interactions between vortices and Dirac fermions in $d-wave$ superconductors turn out to play an important role for the topological order in the perspectives of global one-form symmetries when such Dirac fermions become gapped due to possible ``chiral symmetry breaking".
On the other hand, the existence of massless Dirac fermions gives rise to the fact that low energy fluctuations such as both supercurrent and $U(1)$ gauge-field fluctuations become critical in the large number of fermion flavors.
Interestingly, the discrete one-form global symmetry still remains in the resulting effective field theory of the large$-N_{f}$ limit to be spontaneously broken in the superconducting phase.
In other words,  a spontaneous breaking of a discrete one-form symmetry
may coexist with quantum criticality in $d-wave$ superconductors.

One may criticize that the role of massless Dirac fermions in the notion of topological order for $d-wave$ superconductors is not clarified in a rigorous sense.
The coexistence between the spontaneous breakdown of the discrete one-form symmetry and the quantum criticality should be verified with a concrete mathematical machinery.
One possible way of analysis would be the use of
't~Hooft anomaly,
which is an obstruction of gauging global symmetries \cite{tHooft:1979rat}.
Existence of a 't~Hooft anomaly indicates that the ground state cannot be trivial,
and it provides us with useful information to constrain the phase diagram in a nonperturbative way.
This aspect remains as an interesting future project.

\begin{acknowledgments}
K. K. was supported by the Ministry of Education, Science, and Technology (No. NRF-2015R1C1A1A01051629 and No. 2011-0030046) of the National Research Foundation of Korea (NRF) and by TJ Park Science Fellowship of the POSCO TJ Park Foundation. Y. H. was supported in part by the Korean Ministry of Education, Science and Technology, Gyeongsangbuk-do and Pohang City for Independent Junior Research Groups at the Asia Pacific Center for Theoretical Physics. We appreciate helpful discussions with Yuya Tanizaki.
\end{acknowledgments}

\bibliography{refs}

\end{document}